\documentclass{aastex}
\usepackage{spr-astr-addons}
\usepackage{url}
\usepackage{graphicx}
\usepackage{longtable}
\usepackage{comment}
\usepackage{natbib}
\usepackage{rotating}
\usepackage{verbatim}
\usepackage{morefloats}
\usepackage{lscape}
\usepackage{threeparttable}
\usepackage{wasysym}
\usepackage{txfonts}
\usepackage{gensymb}
\usepackage{multirow}
\usepackage{upgreek}
 \usepackage{booktabs}
\usepackage[verbose]{placeins}
\usepackage{blindtext}
\usepackage{appendix}
\usepackage{enumitem}

\RequirePackage{color}

\newcommand{\emaila}{f.massaro@unito.it}

\newcommand{\bzcat}{{\it Roma-BZCAT}}
\newcommand{\wse}{{\it WISE}}
\newcommand{\chn}{{\it Chandra}}
\newcommand{\swf}{{\it Swift}}
\newcommand{\suz}{{\it Suzaku}}
\newcommand{\fer}{{\it Fermi}}

\begin{document}

\title{The Gamma-ray Blazar Quest: \\ new optical spectra, state of art and future perspectives}
%\slugcomment{}
%% Running heads
\shorttitle{Gamma-ray Blazar Quest}
\shortauthors{F. Massaro et al.}

\author{
F. Massaro\altaffilmark{1,2}, N. \'Alvarez Crespo\altaffilmark{1,2}, R. D'Abrusco\altaffilmark{3}, M. Landoni\altaffilmark{4}, N. Masetti\altaffilmark{5,6}, F. Ricci\altaffilmark{7}, D. Milisavljevic\altaffilmark{8}, A. Paggi\altaffilmark{8}, V. Chavushyan\altaffilmark{9}, E. Jim\'enez-Bail\'on\altaffilmark{10}, V. Pati\~no-\'Alvarez\altaffilmark{9}, J. Strader\altaffilmark{11}, L. Chomiuk\altaffilmark{11}, F. La Franca\altaffilmark{7}, Howard A. Smith\altaffilmark{4} \& G. Tosti\altaffilmark{12}
} 
\email{\emaila}

\altaffiltext{1}{Dipartimento di Fisica, Universit\`a degli Studi di Torino, via Pietro Giuria 1, I-10125 Torino, Italy}
\altaffiltext{2}{Istituto Nazionale di Fisica Nucleare, Sezione di Torino, I-10125 Torino, Italy}
\altaffiltext{3}{Department of Physical Sciences, University of Napoli Federico II, via Cinthia 9, 80126 Napoli, Italy}
\altaffiltext{4}{INAF-Osservatorio Astronomico di Brera, Via Emilio Bianchi 46, I-23807 Merate, Italy}
\altaffiltext{5}{INAF - Istituto di Astrofisica Spaziale e Fisica Cosmica di Bologna, via Gobetti 101, 40129, Bologna, Italy}
\altaffiltext{6}{Departamento de Ciencias F\'{\i}sicas, Universidad Andr\'es Bello, Fern\'andez Concha 700, Las Condes, Santiago, Chile}
\altaffiltext{7}{Dipartimento di Matematica e Fisica, Universit\`a Roma Tre, via della Vasca Navale 84, I-00146, Roma, Italy}
\altaffiltext{8}{Harvard - Smithsonian Center for Astrophysics, 60 Garden Street, Cambridge, MA 02138, USA}
\altaffiltext{9}{Instituto Nacional de Astrof\'{i}sica, \'Optica y Electr\'onica, Apartado Postal 51-216, 72000 Puebla, M\'exico}
\altaffiltext{10}{Instituto de Astronom\'{\i}a, Universidad Nacional Aut\'onoma de M\'exico, Apdo. Postal 877, Ensenada, 22800 Baja California, M\'exico}
\altaffiltext{11}{Department of Physics and Astronomy, Michigan State University, East Lansing, MI 48824, USA}
\altaffiltext{12}{Dipartimento di Fisica, Universit\`a degli Studi di Perugia, 06123 Perugia, Italy}

\begin{abstract}
We recently developed a procedure to recognize $\gamma$-ray blazar candidates within the positional uncertainty regions of the unidentified/unassociated $\gamma$-ray sources (UGSs). Such procedure was based on the discovery that \fer\ blazars show peculiar infrared colors. However, to confirm the real nature of the selected candidates, optical spectroscopic data are necessary. Thus, we performed an extensive archival search for spectra available in the literature in parallel with an optical spectroscopic campaign aimed to reveal and confirm the nature of the selected $\gamma$-ray blazar candidates. Here, we first search for optical spectra of a selected sample of $\gamma$-ray blazar candidates that can be potential counterparts of UGSs using the Sloan Digital Sky Survey (SDSS DR12). This search enables us to update the archival search carried out to date. We also describe the state-of-art and the future perspectives of our campaign to discover previously unknown $\gamma$-ray blazars.
\end{abstract}


\section{Introduction}
\label{sec:intro}
The first observations performed with the Energetic Gamma Ray Experiment Telescope \citep[EGRET;][]{thompson93} revealed that the largest fraction of sources detected in the MeV-GeV energy range were unidentified/unassociated. According to the latest versions of the 3$^{rd}$ EGRET \citep[][]{hartman99,casandjian08} source catalog, the fraction of $\gamma$-ray objects with an unknown origin is $\sim$60\%. This is mostly due to the large positional uncertainties of the $\gamma$-ray sources being up to an order of magnitude greater than those at lower energies. Thus, the association of $\gamma$-ray sources with their low energy counterparts is one of the most demanding task for modern $\gamma$-ray astronomy \citep[see e.g.,][]{thompson08,massaro12a,review}. 

The study of the unidentified/unassociated $\gamma$-ray sources (UGSs) was set as one of the major scientific goals of the \fer\ mission \citep{atwood09}. Thanks to the major improvements achieved in the source localization with the Large Area Telescope (LAT) on board of \fer, the ``source association task'' has been greatly simplified. This was also due to the improvements of statistical methods used to search for low energy counterparts of \fer\ detected objects \citep[see e.g.,][]{abdo10a,abdo10b}. Difficulties on the source association in $\gamma$-rays is partially mitigated by another aspect: the largest fraction of known $\gamma$-ray emitters is associated with the rarest class of active galaxies: the blazars. In the \fer-Large Area Telescope (LAT) third source catalog \citep[3FGL;][]{acero15} blazars constitute about 36\% of the \fer\ sources.

Blazars are generally divided in two classes on the basis of the equivalent width (EW) of their optical spectral features \citep[see e.g.,][]{stickel91}. When presenting featureless spectra or with emission/absorption lines with EW$<$5\AA\ they are classified as BL Lac objects, while those showing quasar-like optical spectra coupled with a flat radio spectrum are known as flat spectrum radio quasars \citep[FSRQ; see also][for more details on the blazar properties]{urry95}. Following the nomenclature proposed in the multifrequency catalog of Blazars \citep[\bzcat;][]{massaro15b}, we indicate the former class as BZB while the latter one as BZQ. We also considered the blazar of galaxy type (indicated as BZG), according to the definition presented in the latest release of the \bzcat\ catalog\footnote{http://www.asdc.asi.it/bzcat/} \citep{massaro09}, being sources whose multifrequency emission exhibits some properties of blazars but appears dominated by the host galaxy contribution in the optical-ultraviolet energy range \citep{massaro12b,massaro14}. 

Across all the releases of the \fer-LAT source catalogs, the fraction of UGSs does not seem to decrease significantly. In the First \fer-LAT source catalog \citep[1FGL][]{abdo10a} there were 630 out of 1451 unassociated \fer\ objects (i.e., $\sim$43\%), fraction that decreases to 34\% in the Second \fer-LAT source catalog \citep[2FGL][]{nolan12} then being $\sim$33\% in the 3FGL. This is in agreement with the fact that even as more distant and/or less luminous sources are found, the UGS fraction remains about the same. This indicates that, even eight years after the launch of \fer, unveiling the UGS nature is still an unsolved issue.

According to the 3FGL analysis the second largest population is constituted by the blazar candidates of uncertain type (BCUs) defined in the 3FGL and in the Third Catalog of Active Galactic Nuclei Detected by the \fer-LAT \citep[3LAC;][see following sections for additional details]{ackermann15a}. This corresponds to a revised definition of the previous $\gamma$-ray source class of active galaxies of uncertain type (AGUs) \citep[see e.g., the First and the Second Catalog of Active Galactic Nuclei Detected by the \fer-LAT, 1LAC and 2LAC, respectively][]{abdo10b,ackermann11}. BCUs are most probably all blazars \citep[see e.g.,][]{massaro12c,crespo16c,chiaro16}, but the lack of spectroscopic information does not permit us to confirm their nature. 

Follow up observations aiming to search for the potential counterparts of the UGSs and to confirm BCUs have been carried out in the radio \citep[e.g.,][]{kovalev09,hovatta12,petrov13,hovatta14,schinzel15}, even below 1GHz \citep{ugs3,ugs6,mwabl}, in the sub millimeter ranges \citep{giommi12,lopez13} and in the X-rays with \swf\ \citep[e.g.,][]{mirabal09,paggi13,takeuchi13,stroh13,acero13} as well as with \chn\ and \suz\ \citep[e.g.,][]{maeda11,cheung12,kataoka12,takahashi12}. Additional archival analyses were also carried out using multifrequency surveys and catalogs \citep[see e.g.,][]{cowperthwaite13,BATcan,blarch}.

Using the \wse\ all-sky survey \citep{wright10}, we showed that in the IR color-color diagrams the $\gamma$-ray blazars, dominated by non-thermal emission, lie in a distinct region well separated from that occupied by other extragalactic sources \citep{paper1,paper2,connect}. On the basis of this discovery, we built new procedures to recognize $\gamma$-ray blazar candidates lying within the positional uncertainty regions of the UGSs  \citep{ugs1,ugs2,wibrals}. 

Then in 2012 we started an optical spectroscopic campaign to confirm the nature of both $\gamma$-ray blazar candidates selected according to our IR procedure, UGS potential counterparts and BCUs associated in the \fer\ catalogs.  All the spectroscopic observations collected during our campaign are already published \citep[see e.g.,][]{paggi14,massaro15c,landoni15,ricci15,crespo16a,crespo16b}. During our campaign we also continuously searched in the optical databases to exclude targets for which an optical spectrum became recently available (see e.g., Massaro et al. 2014, Massaro et al. 2015a, \'Alvarez Crespo et al. 2016c, for the spectra of the Sloan Digital Sky Survey, SDSS, DR12 and of the Six-degree-Field Galaxy Survey, 6dFGS).

In the present paper we aim to present the state-of-art of our {\it hunt} for $\gamma$-ray blazar candidates. Scientific objectives of the current analysis can be summarized as follows:
\begin{enumerate}
\item searching for optical spectra of radio sources that could be potential counterparts of 3FGL UGSs lying in the SDSS footprint;
\item presenting an overview of the results of our optical spectroscopic campaign achieved to date;
\item discussing on the future perspectives towards the preparation of the 4FGL catalog.
\end{enumerate}

The paper is organized as follows. In \S~\ref{sec:new} we described our search for radio sources having a blazar-like optical spectrum and lying within the positional uncertainty region of the UGSs in the footprint of the SDSS. Then \S~\ref{sec:state} and \S~\ref{sec:summary} are devoted to present results and the summary of our optical spectroscopic campaign, respectively. Finally in \S~\ref{sec:future} we speculate on future perspectives towards a better understanding of the unknown $\gamma$-ray sky.

We used cgs units unless stated otherwise through the whole paper, and spectral indices, $\alpha$, are defined by flux density, S$_{\nu}\propto\nu^{-\alpha}$ considering sources with a flat spectrum when $\alpha<$0.5.

\begin{table*}[!ht]
\tiny
\caption{Summary of source details for the eleven blazars found within the sample of UGSs lying in the SDSS footprint (See \S~\ref{sec:new} for more details).}
\label{tab:new}
\begin{tabular}{|lllllllll|}
\hline
3FGL & SDSS & \wse\  & class & $z$ & 1FGL & 2FGL & 1FHL & BZCAT \\
name & name & name  &          &              & name & name & name & name   \\
\hline
\hline 
J0158.6+0102 & J015852.77+010132.9 & J015852.76+010132.9 & bzb & 0.0 &  & J0158.4+0107 &  & 5BZU J0158+0101\\
J0234.2-0629 & J023410.30-062825.7 & J023410.28-062825.8 & bzb & 0.0 &  &  &  \\
J1103.3+5239 & J110249.84+525012.6 & J110249.86+525012.6 & bzq & 0.68984 &  &  &  \\
J1105.7+4427 & J110544.28+442830.4 & J110544.29+442830.6 & bzb & 0.74641 &  &  &  \\
J1129.0+3758 & J112904.78+375844.6 & J112904.78+375844.6 & bzb & 0.0 & J1129.3+3757 & J1129.5+3758 &  \\
J1301.5+3333 & J130129.14+333700.3 & J130129.16+333700.2 & bzq & 1.00826 &  & J1301.6+3331 &  & 5BZQ J1301+3337\\ % 5C 12.170 
J1330.4+5641 & J133040.69+565520.1 & J133040.67+565520.1 & bzb & 0.0 &  &  &  & 5BZB J1330+5655\\
J1411.1+3717 & J141130.56+372245.5 & J141130.51+372246.4 & bzb & 0.0 &  &  &  \\
J1731.9+5428 & J173340.32+542636.8 & J173340.31+542636.7 & bzb & 0.0 &  & J1730.8+5427 &  \\
J2145.5+1007 & J214530.19+100605.4 & J214530.19+100605.5 & bzb & 0.0 &  &  &  \\
J2223.3+0103 & J222329.57+010226.6 & J222329.57+010226.7 & bzb & 0.0 & J2223.3+0103 & J2223.4+0104 & J2223.4+0104 & 5BZB J2223+0102\\ %  NVSS J222329+010226 
\hline 
\end{tabular}\\
Column description: 
 (1) name reported in the Third \fer-LAT source catalog (3FGL);  (2) SDSS name;  (3) \wse\ name;
 (4) Spectroscopic class;  (5) redshift;  (6) name in the First \fer-LAT source catalog (1FGL);
 (7)  name in the Second \fer-LAT source catalog (2FGL);  (8)  name in the First Fermi-LAT Catalog of Sources above 10 GeV (1FHL);  (9) \bzcat\ name.
\end{table*}

\section{Searching for $\gamma$-ray BL Lac candidates in the footprint of the Sloan Digital Sky Survey}
\label{sec:new}

\begin{figure}[!h]
\includegraphics[height=6.4cm,width=8.4cm,angle=0]{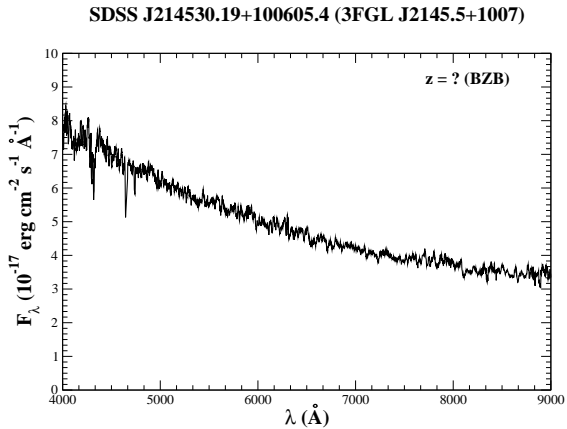}
\caption{The optical spectrum of SDSS J214530.19+100605.4, classified as a BZB at unknown redshift, lying within the positional uncertainty region of the UGS 3FGL J2145.5+1007.} 
\label{fig:spectra}
\end{figure}

As previously stated, our optical spectroscopic campaign was carried out both observing $\gamma$-ray blazar candidates and searching major optical databases and surveys. To perform the literature search we crossmatched the radio/IR position of the $\gamma$-ray blazar candidates selected for the UGSs as well as those of the BCUs reported in the \fer\ catalogs with those reported in the comparison survey as, for example, the SDSS. Thus to complete and update our literature search we decided to carry out the following additional analysis. 

We considered all the UGSs lying in the footprint of the SDSS DR12 \citep{alam15}. Then we searched for all sources having the SDSS spectrum available and with a radio counterpart in the Faint Images of the Radio Sky at Twenty-centimeters (FIRST) survey \citep[see e.g.,][]{becker95,white97}. Finally, we selected only those SDSS sources with a BL Lac or a quasar-like optical spectrum. The latter are considered blazar-like when showing a flat radio spectrum.

This further investigation led to the discovery of six new BL Lacs and one BZQs, potential counterparts of \fer\ UGSs, since one BZQ and two BZBs were recently reported in the latest release of the \bzcat\ but not in the 3FGL. The spectra of one BL Lac is shown in Fig.\ref{fig:spectra} as template of those selected in our analysis. Source details for these eleven blazars are reported in Table~\ref{tab:new} where we report (i) their 3FGL names together with those of the SDSS and the \wse\ potential low energy counterparts; (ii) our classification; (iii) the redshift estimate when possible; (iv) alternative \fer\ names from previous catalogs and (v) the \bzcat\ name.

\begin{figure}[!b]
\includegraphics[height=6.4cm,width=8.4cm,angle=0]{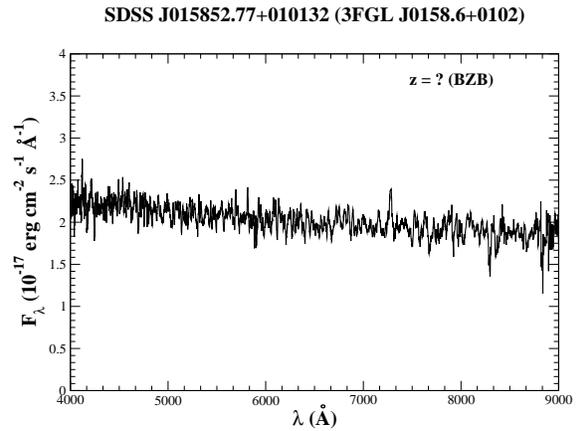}
\caption{The featureless optical spectrum of SDSS J015852.76+010132.9, potential low energy counterpart of the unassociated \fer\ source: 3FGL J0158.6+0102. The source is classified by us as a BL Lac object.} 
\label{fig:0158}
\end{figure}

It is worth noting that the UGS 3FGL J0158.6+0102 could be potentially associated with SDSS J015852.76+010132.9 classified as BZU in the latest release of the \bzcat\ 5BZU J0158+0101 but appearing as a BL Lac object in the SDSS presented here (see Fig.~\ref{fig:0158}). Then we also remark that 3FGL J2223.3+0103 is missing from 3LAC but present in 2FGL class AGU without redshift. 

Finally, we emphasize that even if four blazars are already present in the latest version of the \bzcat\ \citep{massaro15c} there were not indicated as counterparts of 3FGL sources because the latter catalog was released previously.

\section{State-of-art of our optical spectroscopic campaign}
\label{sec:state}
Here we present the summary of the results achieved during spectroscopic campaign considering the latest release of the \fer\ catalog available to date (i.e., 3FGL). Given the differences in the $\gamma$-ray analysis between the releases of the \fer-LAT source catalogs we also discuss some targets that belong to the 1FGL and the 2FGL catalogs only.

We only observed those AGUs that had not been otherwise observed at the time of the 2FGL release \citep[see e.g.,][]{shaw13a,shaw13b}. On the other hand, we recently provided an important contribution to the classification of the 3FGL BCUs obtaining optical spectra for about 1/5 of them \citep[see e.g.,][and references therein]{refined}. 

Regarding the UGSs the impact of our campaign was crucial to decrease their percentage in the 1FGL and in the 2FGL \citep[see also][for a recent review on the sources listed in both these \fer\ catalogs]{refined}, while for the 3FGL the number of newly discovered blazars correspondent to the UGSs is rather small (see following Section). This is mostly due to the fact that the catalog of IR selected blazar-like sources \citep{wibrals} was used to search for associations of the 3FGL sources, thus providing a significant number of BCUs that we preferred to follow up during our spectroscopic campaign.

\subsection{General Results}
All the 3FGL sources observed to date are reported in Table~\ref{tab:main1}, \ref{tab:main2} and \ref{tab:main3} where we provide (i) their \fer\ names together with their (ii) $\gamma$-ray classification, when present; (iii) the associated counterpart for the BCUs and (iv) the infrared (i.e., \wse\ or 2MASS) name of the counterpart observed in our campaign; then we also present (v) our optical classification: BZB for BL Lac objects, BZQ for flat spectrum radio quasars and quasar (QSO) for sources having no radio counterpart and/or not a flat radio spectrum but a quasar-like (i.e., broad emission lines) optical behavior, respectively. Then we also show (vi) the reference among our campaign papers, (vii) the \bzcat\ name. For our optical classification we adopted the \bzcat\ criteria, including the definition of blazar of galaxy type (BZG) for sources having an SED with a significant dominance of the host galaxy emission over the nuclear one \citep[see][for more details]{massaro15b}. It is also worth noting that QSOs, having no blazar-like radio emission, could be contaminants of our selection procedure, as described in D'Abrusco et al. (2014).

The $\gamma$-ray classification reported in Table~\ref{tab:main}, \ref{tab:main2} and \ref{tab:main3} corresponds to the one of the 3LAC \citep{ackermann15a}. In particular we distinguish among BCU of type I, II and III on the basis of the following criteria. These corresponds to the a, b, c subclasses of the 3LAC, respectively.
\begin{enumerate}[label=(\Roman*)]
\item a BZU object (blazar of uncertain/transitional type) in the \bzcat;
\item a source with multiwavelength data in one of the catalogs used for the 3LAC associations, with a flat radio spectrum, and a typical two-humped, blazar-like spectral energy distribution (SED);
\item a source with a radio and an X-ray counterparts also showing a typical two-humped, blazar-like SED.
\end{enumerate}

Since the beginning of our follow up spectroscopic observations in 2012 we pointed 198 unique targets correspondent to 3FGL sources. 

For 3FGL sources we summarize our results as reported below.
\begin{itemize} 
\item We found 75 BZBs all classified in the 3FGL as BL Lacs with the only exception of that associated with 3FGL J2356.0+4037 appeared as a BZG. Most of these targets (i.e., more than $\sim$60\%) were unclassified counterparts of 2FGL sources for which their nature is actually reported in the 3FGL thanks to our campaign. This occurred because a significant fraction of our observations begun between the 2FGL and the 3FGL releases.
\item Nine 3FGL sources associated with flat spectrum radio quasars were also observed, all resulting as BZQ and with a $z$ estimate.
\item We observed 88 BCUs. In this sample we found : 14 BZQs and 10 BZGs all with known redshifts; 1 quasar and 1 source with still uncertain classification plus 62 BZBs, 6 also having a good redshift estimate and 2 with an uncertain $z$ value. 
\item We observed 26 blazar-like objects that could be potential counterparts of \fer\ UGSs, selected with their IR colors. For six UGSs we pointed a quasar, most probably a contaminant of our selection methods unless future radio follow up observations will reveal a flat radio spectrum. The remaining sources are indeed all blazar-like. We found 2 BZQs, one with known redshift, 1 BZG with a certain $z$ estimate and 17 BZBs with a certain $z$ value reported in only four cases.
\end{itemize}

We also observed 10 unique targets that belong to 1FGL, 2FGL and the First Fermi-LAT Catalog of Sources above 10 GeV \citep[1FHL;][]{ackermann13} catalogs but are not in the 3FGL. Results on this sample are thus summarized below.
\begin{itemize} 
\item The potential counterparts of 1FHL J0030.1-1647 and 1FHL J0044.0-1111, both UGS of the 1FHL catalog, could be potentially associated to two newly discovered BZBs at known redshifts, 0.237 and 0.264, respectively \citep{crespo16b}.
\item For the three 1FGL sources we found: 1FGL J0411.6+5459 potentially associated to a new BL Lac at unknown $z$; one QSO contaminant of our selection method for the UGS 1FGL J2110.3+3820; and 1FGL J1422.7+3743 for which we confirmed the BZB classification available in the literature.
\item We observed 3 AGUs of the 2FGL subsample all classified as BZQs and all with a certain redshift estimate provided. Then we also confirm that 2FGL J0819.6-0803 is associated with a BZB at redshift unknown. One additional UGS, 2FGL J1745.6+0203, was also observed searching for blazar-like sources within the \fer\ positional uncertainty region, but we pointed another QSO. 
\end{itemize}

\subsection{Additional source details}
During our campaign we also re-observed a few sources, mostly BL Lacs for which the redshift $z$ estimate was uncertain or unknown, to obtain a better measurement and also to catch some transitional states as occurred in a couple of cases (see \'Alvarez Crespo et al. 2016a,b).  In three cases we also pointed a different target than that actually reported in the 3LAC as described below. Thus we also add the following notes with respect to the results presented in the published papers and previously summarized.

\begin{itemize}
\item In \'Alvarez Crespo et al. (2016a) for the source 3FGL J0653.6+2817 associated with the BCU: GB6 J0653+2816 we pointed a different target resulting as a BZG: WISE J065340.46+281848.5 at uncertain $z>$0.45 instead of WISE J065344.26+281547.5. The same occurred for the BCU associated with 3FGL J1013.5+3440 for which WISE J101336.51+344003.6 was observed instead of OL 318. In addition to them, in Massaro et al. (2015a) for 3FGL J1549.0+6309 the reported ``candidate association'' pointed during our campaign is: NVSS J154828+631050. The source associated in the 3FGL/3LAC catalog has the coordinates of SDSS J154957.32+631007.3 (a.k.a. CRATES J1549+6310) but the counterpart name seems incorrect being: SDSS J154958.45+631021.2; in both cases out pointed target differs from them.

\item The BCU NVSS J132210+084231 associated with 3FGL J1322.3+0839 was observed in \'Alvarez Crespo et al. (2016a) and \'Alvarez Crespo et al. (2016c) in a different state, since it appeared as a BZB first and a BZQ at $z=$0.32549 later. The optical spectrum of 3FGL J0158.6+0102 is reported Massaro et al. (2014) as well as in the current paper. This source showed a BZQ spectrum in the past not confirmed by the most recent observation. Thus these two sources can be considered as transitional sources present in our sample in addition to those discovered in (\'Alvarez Crespo et al. 2016a,c): 3FGL J0433.1+3228 and 3FGL J1040.4+0615. These two \fer\ objects were originally classified as BZQ while had a typical BZB optical spectrum when pointed by us.

\item Only for the AGU: TXS 1801+253 associated in the 2FGL with 2FGL J1803.6+2523c, we were not able to find a \wse\ counterpart, due to a confusion with a nearby star \citep[see][for more details]{crespo16c}.

\item We re-pointed the associated counterparts of 3FGL J1340.6-0408, 3FGL J1314.8+2349, 3FGL J2323.9+4211, 3FGL J1129.0+3758 and 3FGL J2223.3+0103 during our campaign but they always appeared as BZBs at unknown redshift.

\item We found that the associated counterparts of 3FGL J0522.9-3628 is a BZQ while those of 3FGL J0009.6-3211 and 3FGL J2356.0+4037 are indeed BZGs at redshifts 0.002542 and 0.131, respectively.
\end{itemize}

\section{Summary and Conclusions}
\label{sec:summary}
The \wse\ all-sky survey \citep{wright10} allowed us to distinguish $\gamma$-ray emitting blazars from other extragalactic sources. On the basis of the mid-IR colors we developed a set of procedures to select potential counterparts of the UGSs and to recognize blazar candidates within the AGUs and the BCUs listed in the \fer\ catalogs. Motivated by preliminary results of our IR selection, we carried out an optical spectroscopic campaign aiming to verify the blazar-like nature of UGS potential counterparts As secondary goal we also obtained a classification for AGUs and BCUs present in the \fer\ catalogs.

During our campaign we collected and analyzed 223 unique spectra and the major results achieved are summarized as follows.

\begin{enumerate}
\item The largest fraction of pointed targets are classified as BZBs (i.e., 173/223). This strongly suggests that \fer\ survey is extremely useful to discover new BZBs and confirms that these are one of most elusive class of active galaxies.

\item We obtained 49 certain redshifts for the blazars observed thanks to the presence of emission/absorption lines or only absorption features due to intervening systems.

\item We discovered a handful of transitional sources: blazars with a different spectrum available in the literature and thus classified differently.

\item During our campaign we found two BL Lac objects without radio counterparts in the major radio surveys \citep{paggi14,ricci15}. This discovery could open new scenarios on the blazar phenomenon and could potentially make the $\gamma$-ray association task more challenging since a large fraction of the \fer\ associations come from radio surveys/catalogs.

\item Within the sample of newly discovered BL Lacs with a $z$ estimate, presented in Section~\ref{sec:new}, it is worth noting that one of them lie above redshift $z=$0.7 (this occurs only for 11 BL Lacs from the \bzcat\ listed in the \fer\ catalogs).
\end{enumerate}

Given the small number of QSOs found during our campaign (i.e., $\sim$4\%) in addition to a handful of unpublished spectra that did not have enough signal-to-noise ratio to clearly classify the observed targets, our results strongly supports the reliability of our IR selection method \citep[see][for more details]{dabrusco16}.

Finally, we emphasize that results obtained thanks to our campaign were used to classify the sources listed in the latest release of the \fer\ catalogs \citep[i.e., 3FGL, 3LAC and 2FHL][]{ackermann16} as well as to increase those in the \bzcat.

\section{Future perspectives}
\label{sec:future}
Our spectroscopic follow up campaign at optical wavelengths is still on-going and we expect to have an additional $\sim$80 targets presented by the end of 2016. 

These optical observations of $\gamma$-ray blazar candidates are crucial to build the luminosity function of BL Lacs \citep[see e.g.,][]{ajello12,ajello14}, confirm potential targets for the Cherenkov Telescope Array (CTA) selected on the basis of multifrequency properties \citep[see e.g.,][]{tev,arsioli15}, obtain stringent limits on the dark matter annihilation in sub-halos \citep[see e.g.,][]{zechlin12,berlin14} and/or from the estimates of the extragalactic $\gamma$-ray background \citep[EGB][]{ajello15}, search for counterparts of new flaring $\gamma$-ray sources \citep[see e.g.][]{bernieri13} and identify potential counterparts of new \fer\ sources discovered with detection algorithms different from those used to build the \fer\ catalog \citep[see e.g.,][]{campana15,campana16a,campana16b}. Studies on the UGSs are also crucial to discover new BL Lacs that might improve our knowledge on their contribution to the extragalactic $\gamma$-ray background \citep[EGB; see e.g.,][]{ajello14}. The BL Lac contribution considered together with that arising from other $\gamma$-ray emitters as for example radiogalaxies \citep[see e.g.][]{inoue11,lobes,dimauro14} and/or star forming galaxies \citep[see e.g.,][]{ackermann12a} could help searching for a potential dark matter signature in the EGB spectrum (Ackermann et al. 2015a, Di Mauro \& Donato 2015, Fermi Collaboration 2015). Moreover optical spectroscopic campaign are also useful to obtain new redshifts necessary to estimate the imprint of the extragalactic background light in their spectra \citep[see e.g.,][]{ackermann12b}.

At the current moment the major problem we foresee to continue after 2016 resides in the strategy originally developed, even if it has been successful to date. To minimize the impact on telescope schedules and maximize the scientific return, we proposed a small subsample of targets each time. Consequently, each publication presents only a limited amount of spectra, in the range between 20 and 50. However, it is worth noting that even if the number of the \fer\ sources for each catalog is growing consistently with the expectations: 1451 in the 1FGL, 1873 in the 2FGL and now 3033 in the 3FGL, the fraction of UGSs is rather constant. Thus if the expected number of sources in the next \fer\ catalog (i.e., 4FGL) will be $\sim$4500 sources, the number of UGSs and BCUs expected of the order of 2000. Consequently, assuming we can still obtain observing nights at the telescopes used to date, 
collecting $\sim$200 spectra in about 3 years, it will be impossible to achieve our final goal of ``resolving'' the $\gamma$-ray sky on a few years time scale.

To complete our campaign, considering the sources listed in the 3FGL, the number of targets that must be observed is approximately 550, mostly concentrated in the Southern hemisphere, and correspondent to $\sim$40 observing nights, assuming to use 4 meter class telescopes/facilities. Thus, the only strategy suitable to achieve our final goal is a survey/large program. Completing these observations will allow us to assign a certain classification to all the BCUs listed in the 3FGL and decrease the fraction of 3FGL UGSs by a factor of 5-10\%, assuming results similar to those achieved to date.

\acknowledgements
We thank the anonymous referee for useful comments that led to improvements in the paper.
% Friends
% Montalcini
F.M. gratefully acknowledges the financial support of the Programma
Giovani Ricercatori -- Rita Levi Montalcini -- Rientro dei Cervelli (2012) awarded by the Italian Ministry of Education, Universities and Research (MIUR).
% grants
This investigation is supported by the NASA grants NNX12AO97G and NNX13AP20G.
The work by G. Tosti is supported by the ASI/INAF contract I/005/12/0.
H. A. Smith acknowledges partial support from NASA/JPL grant NNX14AJ61G.
H. Ot\'i-Floranes is funded by a postdoctoral UNAM grant.
V. Chavushyan acknowledges funding by CONACyT research grant 151494 (M\'exico).
% OAN San Pedro Martir Observatory
We thank the staff at the Observatorio Astron\'omico Nacional in San Pedro M\'artir (M\'exico) for all their help during the observation runs.
% Loiano
We thank the staff of the Astronomical Observatory of Bologna in Loiano for their assistance during the observations.
% ASDC
Part of this work is based on archival data, software or on-line services provided by the ASI Science Data Center.
% HEASARC
This research has made use of data obtained from the high-energy Astrophysics Science Archive
Research Center (HEASARC) provided by NASA's Goddard Space Flight Center; 
% SIMBAD and NED
the SIMBAD database operated at CDS,
Strasbourg, France; the NASA/IPAC Extragalactic Database
(NED) operated by the Jet Propulsion Laboratory, California
Institute of Technology, under contract with the National Aeronautics and Space Administration.
% WISE
This publication makes use of data products from the Wide-field Infrared Survey Explorer, 
which is a joint project of the University of California, Los Angeles, and 
the Jet Propulsion Laboratory/California Institute of Technology, 
funded by the National Aeronautics and Space Administration.
% 2MASS
This publication makes use of data products from the Two Micron All Sky Survey, which is a joint project of the University of 
Massachusetts and the Infrared Processing and Analysis Center/California Institute of Technology, funded by the National Aeronautics 
and Space Administration and the National Science Foundation.
% SDSS
Funding for SDSS-III has been provided by the Alfred P. Sloan Foundation, the Participating Institutions, the National Science Foundation, and the U.S. Department of Energy Office of Science. The SDSS-III web site is http://www.sdss3.org/. SDSS-III is managed by the Astrophysical Research Consortium for the Participating Institutions of the SDSS-III Collaboration including the University of Arizona, the Brazilian Participation Group, Brookhaven National Laboratory, Carnegie Mellon University, University of Florida, the French Participation Group, the German Participation Group, Harvard University, the Instituto de Astrofisica de Canarias, the Michigan State/Notre Dame/JINA Participation Group, Johns Hopkins University, Lawrence Berkeley National Laboratory, Max Planck Institute for Astrophysics, Max Planck Institute for Extraterrestrial Physics, New Mexico State University, New York University, Ohio State University, Pennsylvania State University, University of Portsmouth, Princeton University, the Spanish Participation Group, University of Tokyo, University of Utah, Vanderbilt University, University of Virginia, University of Washington, and Yale University.
%USNO
This research has made use of the USNOFS Image and Catalogue Archive
operated by the United States Naval Observatory, Flagstaff Station
(http://www.nofs.navy.mil/data/fchpix/).
% TOPCAT
TOPCAT\footnote{\underline{http://www.star.bris.ac.uk/$\sim$mbt/topcat/}} 
\citep{taylor05} for the preparation and manipulation of the tabular data and the images.
% ALADIN
The Aladin Java applet\footnote{\underline{http://aladin.u-strasbg.fr/aladin.gml}}
was used to create the finding charts reported in this paper \citep{bonnarel00}. 
It can be started from the CDS (Strasbourg - France), from the CFA (Harvard - USA), from the ADAC (Tokyo - Japan), 
from the IUCAA (Pune - India), from the UKADC (Cambridge - UK), or from the CADC (Victoria - Canada).

\begin{table*}
\tiny
 \caption{Summary of the targets observed during our optical spectroscopic campaign for the 3FGL sources. (R.A. 00-08)}
\label{tab:main1}
\begin{tabular}{|llllllll|}
\hline
3FGL    & 3FGL   & 3FGL & counterpart & class & $z$ & reference & \bzcat\ \\
name & class & counterpart & name          &           &        &                 & name \\
\hline
\hline 
J0003.8-1151 & bcu II & PMN J0004-1148 & J000404.91-114858.3 & bzb & 0.0 & Crespo+16c & 5BZB J0004-1148\\
J0009.6-3211 & bcu I & IC 1531 & J000935.55-321636.8 & bzg & 0.02542 & Crespo+16c & \\
J0015.7+5552 & bcu II & GB6 J0015+5551 & J001540.13+555144.7 & bzb & 0.0 & Crespo+16a & \\
J0028.8+1951 & bcu III & TXS 0025+197 & J002829.81+200026.7 & bzq & 1.5517 & Crespo+16c & \\
J0030.2-1646 & bcu II & 1RXS J003019.6-164723 & J003019.40-164711.7 & bzb & 0.0 & Crespo+16c & \\
J0043.5-0444 & bcu II & 1RXS J004333.7-044257 & J004334.12-044300.6 & bzb & 0.0 & Crespo+16c & \\
J0103.4+5336 & bll & 1RXS J010325.9+533721 & J010325.89+533713.4 & bzb & 0.0 & Crespo+16a & \\
J0103.7+1323 & bcu III & NVSS J010345+132346 & J010345.74+132345.3 & bzb & 0.49 & Crespo+16b & \\
J0105.3+3928 & bll & GB6 J0105+3928 & J010509.19+392815.1 & bzb & 0.44 & Crespo+16b & 5BZB J0105+3928\\
J0109.1+1816 & bll & MG1 J010908+1816 & J010908.17+181607.5 & bzb & 0.0 & Massaro+14 & 5BZB J0109+1816\\
J0116.3-6153 & bll & SUMSS J011619-615343 & J011619.62-615343.4 & bzb & 0.0 & Landoni+15 & 5BZB J0116-6153\\
J0133.0-4413 & bll & SUMSS J013306-441422 & J013306.37-441421.4 & bzb & 0.0 & Landoni+15 & 5BZB J0133-4414\\
J0143.7-5845 & bll & SUMSS J014347-584550 & J014347.41-584551.4 & bzb & 0.0 & Landoni+15 & 5BZB J0143-5845\\
J0147.0-5204 & bcu I & PKS 0144-522 & J014648.58-520233.5 & bzg & 0.098 & Crespo+16c & 5BZG J0146-5202\\
J0148.3+5200 & bcu III & GB6 J0148+5202 & J014820.33+520204.9 & bzb & 0.0 & Crespo+16a & \\
J0145.6+8600 & bcu III & NVSS J014929+860114 & J014935.28+860115.4 & bzg & 0.15 & Crespo+16a & \\
J0156.9-4742 & bcu II & 2MASS J01564603-4744174 & J015646.03-474417.3 & bzb & 0.0 & Crespo+16c & \\
J0157.0-5301 & bll & 1RXS J015658.6-530208 & J015658.00-530200.0 & bzb & 0.0 & Landoni+15 & 5BZB J0156-5302\\
J0158.6-3931 & bll & PMN J0158-3932 & J015838.10-393203.8 & bzb & 0.0 & Landoni+15 & 5BZB J0158-3932\\
J0158.6+0102 &  &  & J015852.76+010132.9 & bzb & 0.0 & Massaro+16 & 5BZU J0158+0101\\
J0158.6+0102 &  &  & J015852.76+010132.9 & bzq & 1.61? & Massaro+14 & 5BZU J0158+0101\\
J0219.0+2440 & bcu II & 87GB 021610.9+243205 & J021900.40+244520.6 & bzb & 0.0 & Crespo+16a & \\
J0221.2+2518 &  &  & J022051.24+250927.6 & qso & 0.4818 & Paggi+14 & 5BZU J0220+2509\\
J0222.6+4301 & bll & 3C66A & J022239.60+430207.8 & bzb & 0.0 & Paggi+14 & 5BZB J0222+4302\\
J0234.2-0629 &  &  & J023410.28-062825.8 & bzb & 0.0 & Massaro+16 & \\
J0237.5-3603 & bll & RBS 0334 & J023734.04-360328.4 & bzb & 0.0 & Landoni+15 & 5BZB J0237-3603\\
J0238.4-3117 & bll & 1RXS J023832.6-311658 & J023832.48-311658.0 & bzb & 0.232 & Landoni+15 & 5BZB J0238-3116\\
J0253.5+3216 & bcu II & MG3 J025334+3217 & J025333.64+321720.8 & bzq & 0.859 & Ricci+15 & \\
J0255.8+0532 & bcu II & PMN J0255+0533 & J025549.51+053355.0 & bzb & 0.0 & Crespo+16c & \\
J0301.8-7157 & bcu II & PKS 0301-721 & J030138.47-715634.5 & bzq & 0.8232 & Crespo+16c & 5BZQ J0301-7156\\
J0307.3+4916 &  &  & J030727.21+491510.6 & bzb & 0.0 & Crespo+16b & \\
J0309.5-0749 & bll & NVSS J030943-074427 & J030943.23-074427.5 & bzb & 0.0 & Massaro+15a & 5BZB J0309-0744\\
J0316.2-6436 & bll & SUMSS J031614-643732 & J031614.34-643731.4 & bzb & 0.0 & Landoni+15 & 5BZB J0316-6437\\
J0323.6-0109 & bll & 1RXS J032342.6-011131 & J032343.62-011146.1 & bzb & 0.0 & Massaro+14 & 5BZB J0323-0111\\
J0333.6+2916 & bll & TXS 0330+291 & J033349.00+291631.5 & bzb & 0.0 & Crespo+16a & 5BZB J0333+2916\\
J0334.3-4008 & bll & PKS 0332-403 & J033413.65-400825.4 & bzb & 0.0 & Landoni+15 & 5BZB J0334-4008\\
J0335.3-4459 & bll & 1RXS J033514.5-445929 & J033513.88-445943.8 & bzb & 0.0 & Landoni+15 & 5BZB J0335-4459\\
J0339.2-1738 & bcu I & PKS 0336-177 & J033913.70-173600.8 & bzg & 0.06556 & Crespo+16c & \\
J0343.3+3622 & bcu I & OE 367 & J034328.94+362212.4 & bzq & 1.485 & Crespo+16c & 5BZQ J0343+3622\\
J0352.9+5655 & bcu II & GB6 J0353+5654 & J035309.54+565430.7 & bzb & 0.0 & Crespo+16b & \\
J0409.8-0358 & bll & NVSS J040946-040003 & J040946.58-040003.5 & bzb & 0.0 & Massaro+15a & 5BZB J0409-0400\\
J0415.7-4351 &  &  & J041605.82-435514.6 & qso & 0.398 & Landoni+15 & 5BZU J0416-4355\\
J0425.0-5331 & bll & PMN J0425-5331 & J042504.27-533158.2 & bzb & 0.0 & Landoni+15 & 5BZB J0425-5331\\
J0428.6-3756 & bll & PKS 0426-380 & J042840.41-375619.3 & bzb & 1.02? & Landoni+15 & 5BZB J0428-3756\\
J0433.1+3228 & bcu II & NVSS J043307+322840 & J043307.54+322840.7 & bzb & 0.0 & Crespo+16a & \\
J0433.7-6028 & bcu II & PKS 0432-606 & J043334.59-603010.3 & bzq & 0.9301 & Crespo+16c & 5BZQ J0433-6030\\
J0434.4-2341 & bcu I & PMN J0434-2342 & J043428.98-234205.3 & bzb & 0.0 & Crespo+16c & 5BZB J0434-2342\\
J0439.9-1859 & bcu II & PMN J0439-1900 & J043949.72-190101.5 & bzb & 0.0 & Crespo+16c & \\
J0505.9+6114 & bll & NVSS J050558+611336 & J050558.78+611335.9 & bzb & 0.0 & Paggi+14 & 5BZB J0505+6113\\
J0508.2-1936 & bcu II & PMN J0508-1936 & J050818.99-193555.7 & bzq & 1.88 & Crespo+16b & \\
J0521.7+0103 & bcu II & NVSS J052140+010257 & J052140.82+010255.5 & ? & 0.0 & Crespo+16c & \\
J0522.9-3628 & bcu I & PKS 0521-36 & J052257.98-362730.8 & bzq & 0.05655 & Crespo+16c & 5BZU J0522-3627\\
J0537.4-5717 & bll & SUMSS J053748-571828 & J053748.96-571830.1 & bzb & 1.18? & Landoni+15 & 5BZB J0537-5718\\
J0556.0-4353 & bll & SUMSS J055618-435146 & J055618.74-435146.0 & bzb & 0.0 & Landoni+15 & 5BZB J0556-4351\\
J0601.0+3837 & bll & PKS 0601-70 & J060102.86+383829.2 & bzb & 0.0 & Paggi+14 & 5BZB J0601+3838\\
J0604.1-4817 & bll & 1ES 0602-482 & J060408.61-481725.1 & bzb & 0.0 & Landoni+15 & 5BZB J0604-4817\\
J0607.4+4739 & bll & TXS 0603+476 & J060723.25+473947.0 & bzb & 0.0 & Massaro+15a & 5BZB J0607+4739\\
J0609.4-0248 & bll & NVSS J060915-024754 & J060915.06-024754.5 & bzb & 0.0 & Massaro+15a & 5BZB J0609-0247\\
J0612.8+4122 & bll & B3 0609+413 & J061251.18+412237.4 & bzb & 0.0 & Massaro+15a & 5BZB J0612+4122\\
J0618.2-2429 & bcu II & PMN J0618-2426 & J061822.65-242637.7 & bzq & 0.2995 & Crespo+16b & \\
J0644.6+6035 &  &  & J064459.38+603131.7 & bzb & 0.3582 & Paggi+14 & \\
J0650.7+2503 & bll & 1ES 0647+250 & J065046.48+250259.5 & bzb & 0.0 & Massaro+15a & 5BZB J0650+2502\\
J0653.6+2817 & bcu II & GB6 J0653+2816 & J065340.46+281848.5 & bzg & 0.45? & Crespo+16a & \\
J0700.0+1709 & bcu II & TXS 0657+172 & J070001.49+170921.9 & bzq & 1.08 & Crespo+16b & \\
J0700.2+1304 & bcu II & GB6 J0700+1304 & J070014.31+130424.4 & bzb & 0.0 & Crespo+16a & \\
J0708.9+2239 & bcu II & GB6 J0708+2241 & J070858.28+224135.4 & bzb & 0.0 & Massaro+15a & 5BZB J0708+2241\\
J0720.0-4010 & bcu II & 1RXS J071939.2-401153 & J071939.18-401147.4 & bzb & 0.0 & Crespo+16b & \\
J0721.5-0221 &  &  & J072113.90-022055.0 & bzb & 0.0 & Crespo+16b & \\
J0723.2-0728 & bcu III & 1RXS J072259.5-073131 & J072259.68-073134.7 & bzb & 0.0 & Crespo+16c & 5BZB J0722-0731\\
J0728.0+4828 & bcu II & GB6 J0727+4827 & J072759.84+482720.3 & bzb & 0.0 & Crespo+16a & \\
J0730.5-6606 & bcu II & PMN J0730-6602 & J073049.51-660218.9 & bzb & 0.0 & Crespo+16c & \\
J0746.6-4756 & bcu II & PMN J0746-4755 & J074642.31-475455.0 & bzb & 0.0 & Ricci+15 & \\
J0754.8+4824 & bll & GB1 0751+485 & J075445.66+482350.7 & bzb & 0.0 & Massaro+14 & 5BZB J0754+4823\\
J0813.3+6509 & bcu II & GB6 J0812+6508 & J081240.84+650911.1 & bzb & 0.0 & Massaro+15a & 5BZB J0812+6509\\
J0814.1-1012 & bll & NVSS J081411-101208 & J081411.69-101210.2 & bzb & 0.0 & Crespo+16a & 5BZB J0814-1012\\
J0814.7+6428 & bll & GB6 J0814+6431 & J081439.19+643122.0 & bzb & 0.0 & Massaro+15a & 5BZB J0814+6431\\
J0822.9+4041 & fsrq & B3 0819+408 & J082257.55+404149.7 & bzq & 0.8655 & Massaro+14 & 5BZQ J0822+4041\\
J0827.2-0711 & bcu I & PMN J0827-0708 & J082706.16-070845.9 & bzb & 0.0 & Crespo+16c & \\
J0828.8-2420 & bcu III & NVSS J082841-241850 & J082841.74-241851.1 & bzb & 0.0 & Crespo+16b & \\
J0835.4+0930 & bll & GB6 J0835+0936 & J083543.21+093717.9 & bzb & 0.35? & Massaro+14 & 5BZB J0835+0937\\
J0845.1-5458 & bcu II & PMN J0845-5458 & J084502.48-545808.4 & bzb & 0.0 & Ricci+15 & \\
J0849.1+6607 & bll & GB6 J0848+6605 & J084854.60+660609.3 & bzb & 0.0 & Massaro+15a & 5BZB J0848+6606\\
J0855.4+7142 &  &  & J085654.85+714623.8 & bzq & 0.542 & Ricci+15 & \\
\hline 
\end{tabular}\\
Column description: 
 (1) 3FGL name reported; 
 (2) 3FGL $\gamma$-ray classification. Empty field implies that the sources is an UGS;  
 (3) 3FGL associated counterpart if present;
 (4) \wse\ name of the counterpart observed during our optical spectroscopic campaign;
 (5) optical classification
 (6) redshift; question marks indicate uncertain values or redshifts estimated via intervening systems while empty field is used for unknown $z$; 
 (7) reference within the papers published with the results of our campaign;
 (8) the name reported in the \bzcat.\\
Notes: Sources pointed twice are reported in different lines within the table.
\end{table*}

\begin{table*}
\tiny
 \caption{Summary of the targets observed during our optical spectroscopic campaign for the 3FGL sources. (R.A. 08-16)}
\label{tab:main2}
\begin{tabular}{|llllllll|}
\hline
3FGL    & 3FGL   & 3FGL & counterpart & class & $z$ & reference & \bzcat\ \\
name & class & counterpart & name          &           &        &                 & name \\
\hline
\hline 
J0900.0+6754 &  &  & J090038.69+674223.3 & bzb & 0.0 & Massaro+15a & 5BZB J0900+6742\\
J0904.3+4240 & bcu II & S4 0900+42 & J090415.62+423804.5 & bzq & 1.34246 & Crespo+16c & \\
J0917.3-0344 & bcu II & NVSS J091714-034315 & J091714.61-034314.2 & bzg & 0.308 & Crespo+16b & \\
J0921.0-2258 & bcu II & NVSS J092057-225721 & J092057.47-225721.5 & bzb & 0.0 & Crespo+16c & \\
J0924.0+2816 & fsrq & B2 0920+28 & J092351.52+281525.1 & bzq & 0.7442 & Massaro+14 & 5BZQ J0923+2815\\
J0942.1-0756 & bll & PMN J0942-0800 & J094221.46-075953.0 & bzb & 0.0 & Crespo+16a & 5BZB J0942-0759\\
J0950.1+4554 & bll & RX J0950.2+4553 & J095011.82+455320.0 & bzb & 0.3994 & Massaro+14 & 5BZB J0950+4553\\
J1003.6+2608 & bcu I & PKS 1000+26 & J100342.22+260512.8 & bzb & 0.9295 & Crespo+16c & \\
J1007.9+0621 & bll & MG1 J100800+0621 & J100800.81+062121.2 & bzb & 0.6495? & Paggi+14 & 5BZB J1008+0621\\
J1013.5+3440 & fsrq & OL 318 & J101336.51+344003.6 & qso & 0.208 & Crespo+16a & \\
J1016.1+5555 & bcu II & TXS 1012+560 & J101544.43+555100.6 & bzq & 0.677 & Massaro+15a & 5BZQ J1015+5551\\
J1018.3+3542 & fsrq & B2 1015+35B & J101810.97+354239.4 & bzq & 1.228 & Massaro+14 & 5BZQ J1018+3542\\
J1023.1+3952 & fsrq & 4C +40.25 & J102333.50+395312.7 & bzq & 1.3328 & Massaro+14 & 5BZQ J1023+3953\\
J1038.9-5311 & bcu II & MRC 1036-529 & J103840.66-531142.9 & bzq & 1.45 & Crespo+16b & \\
J1040.4+0615 & bcu II & GB6 J1040+0617 & J104031.62+061721.7 & bzb & 0.0 & Crespo+16c & \\
J1040.8+1342 & bcu II & 1RXS J104057.7+134216 & J104057.69+134211.7 & bzb & 0.0 & Crespo+16c & \\
J1042.0-0557 & bcu II & PMN J1042-0558 & J104204.30-055816.5 & bzg & 0.39 & Crespo+16b & \\
J1049.7+1548 &  &  & J104939.34+154837.8 & bzb & 0.3271 & Paggi+14 & 5BZB J1049+1548\\
J1100.5+4020 & bll & RX J1100.3+4019 & J110021.05+401928.0 & bzb & 0.0 & Massaro+14 & 5BZB J1100+4019\\
J1103.3+5239 &  &  & J110249.86+525012.6 & bzq & 0.68984 & Massaro+16 & \\
J1103.9-5357 & bll & PKS 1101-536 & J110352.22-535700.7 & bzb & 0.461? & Landoni+15 & 5BZB J1103-5357\\
J1105.7+4427 &  &  & J110544.29+442830.6 & bzb & 0.74641 & Massaro+16 & \\
J1123.3-2529 &  &  & J112325.37-252857.0 & qso & 0.148 & Ricci+15 & \\
J1125.0-2101 & bcu II & PMN J1125-2100 & J112508.62-210105.9 & bzb & 0.0 & Crespo+16c & \\
J1129.0+3758 &  &  & J112903.24+375656.7 & bzb & 0.0 & Massaro+14 & 5BZB J1129+3756\\
J1129.0+3758 &  &  & J112904.78+375844.6 & bzb & 0.0 & Massaro+16 & \\
J1129.4-4215 & bcu I & SUMSS J113006-421441 & J113007.04-421440.9 & bzb & 0.0 & Crespo+16c & \\
J1141.6-1406 & bcu II & 1RXS J114142.2-140757 & J114141.80-140754.6 & bzb & 0.0 & Ricci+15 & \\
J1200.8+1228 & bcu II & GB6 J1200+1230 & J120040.03+123103.2 & bzb & 0.0 & Crespo+16c & \\
J1209.4+4119 & bll & B3 1206+416 & J120922.78+411941.3 & bzb & 0.0 & Massaro+14 & 5BZB J1209+4119\\
J1217.8+3007 & bll & 1ES 1215+303 & J121752.08+300700.6 & bzb & 0.0 & Ricci+15 & 5BZB J1217+3007\\
J1221.5-0632 &  &  & J122127.20-062847.8 & qso & 0.44 & Crespo+16b & \\
J1222.4+0414 & fsrq & 4C +04.42 & J122222.54+041315.7 & bzq & 0.9642 & Massaro+14 & 5BZQ J1222+0413\\
J1222.7+7952 &  &  & J122358.06+795328.2 & bzb & 0.0 & Massaro+15a & 5BZB J1223+7953\\
J1230.3+2519 & bll & ON 246 & J123014.08+251807.1 & bzb & 0.0 & Massaro+15a & 5BZB J1230+2518\\
J1238.2-1958 & bcu II & PMN J1238-1959 & J123824.39-195913.8 & bzb & 0.0 & Ricci+15 & 5BZU J1238-1959\\
J1243.1+3627 & bll & Ton 116 & J124312.73+362743.9 & bzb & 0.0 & Massaro+14 & 5BZB J1243+3627\\
J1256.3-1146 & bcu I & PMN J1256-1146 & J125615.95-114637.3 & bzg & 0.0579 & Crespo+16c & 5BZG J1256-1146\\
J1259.8-3749 & bcu II & NVSS J125949-374856 & J125949.80-374858.1 & bzb & 0.0 & Ricci+15 & \\
J1301.5+3333 &  &  & J130129.16+333700.2 & bzq & 1.0084 & Massaro+14 & 5BZQ J1301+3337\\
J1301.5+3333 &  &  & J130129.16+333700.2 & bzq & 1.00826 & Massaro+16 & 5BZQ J1301+3337\\
J1311.0+0036 & bll & RX J1311.1+0035 & J131106.47+003510.0 & bzb & 0.0 & Massaro+14 & 5BZB J1311+0035\\
J1314.8+2349 & bll & TXS 1312+240 & J131443.80+234826.7 & bzb & 0.0 & Massaro+14 & 5BZB J1314+2348\\
J1314.8+2349 & bll & TXS 1312+240 & J131443.80+234826.7 & bzb & 0.0 & Paggi+14 & 5BZB J1314+2348\\
J1315.4+1130 & bcu II & 1RXS J131531.9+113327 & J131532.62+113331.7 & bzb & 0.73 & Crespo+16c & \\
J1322.3+0839 & bcu II & NVSS J132210+084231 & J132210.17+084232.9 & bzb & 0.0 & Crespo+16a & \\
J1322.3+0839 & bcu II & NVSS J132210+084231 & J132210.17+084232.9 & bzq & 0.32549 & Crespo+16c & \\
J1328.5-4728 & bcu II & 1WGA J1328.6-4727 & J132840.61-472749.2 & bzb & 0.0 & Ricci+15 & \\
J1330.6+7002 & bll & NVSS J133025+700141 & J133025.82+700138.7 & bzb & 0.0 & Massaro+15a & 5BZB J1330+7001\\
J1330.4+5641 &  &  & J133040.67+565520.1 & bzb & 0.0 & Massaro+16 & 5BZB J1330+5655\\
J1331.8+4718 & fsrq & B3 1330+476 & J133245.24+472222.6 & bzq & 0.6687 & Massaro+14 & 5BZQ J1332+4722\\
J1340.6-0408 & bcu II & NVSS J134042-041006 & J134042.02-041006.8 & bzb & 0.0 & Ricci+15 & \\
J1340.6-0408 & bcu II & NVSS J134042-041006 & J134042.02-041006.8 & bzb & 0.0 & Crespo+16c & \\
J1342.7+0945 & bcu II & NVSS J134240+094752 & J134240.02+094752.4 & qso & 0.28279 & Crespo+16c & \\
J1346.9-2958 & bcu II & NVSS J134706-295840 & J134706.88-295842.4 & bzb & 0.0 & Ricci+15 & \\
J1351.4+1115 & bll & RX J1351.3+1115 & J135120.84+111453.0 & bzb & 0.0 & Massaro+14 & 5BZB J1351+1114\\
J1359.9-3746 & bll & PMN J1359-3746 & J135949.71-374600.7 & bzb & 0.0 & Ricci+15 & 5BZB J1359-3746\\
J1406.0-2508 & bcu II & NVSS J140609-250808 & J140609.60-250809.2 & bzb & 0.0 & Ricci+15 & \\
J1410.4+2821 & bll & RX J1410.4+2821 & J141029.56+282055.6 & bzb & 0.0 & Massaro+14 & 5BZB J1410+2820\\
J1411.1+3717 &  &  & J141130.51+372246.4 & bzb & 0.0 & Massaro+16 & \\
J1412.0+5249 & bcu I & SBS 1410+530 & J141149.44+524900.2 & bzg & 0.07649 & Crespo+16c & \\
J1434.6+6640 & bcu II & 1RXS J143442.0+664031 & J143441.46+664026.5 & bzb & 0.0 & Crespo+16a & \\
J1442.0+4348 & bll & SDSS J144207.15+434836.6 & J144207.15+434836.7 & bzb & 0.0 & Massaro+14 & 5BZB J1442+4348\\
J1444.0-3907 & bll & PKS 1440-389 & J144357.20-390840.0 & bzb & 0.0 & Landoni+15 & 5BZB J1443-3908\\
J1511.8-0513 & bcu III & NVSS J151148-051345 & J151148.56-051346.9 & bzb & 0.0 & Crespo+16a & \\
J1521.8+4340 & fsrq & B3 1520+437 & J152149.61+433639.2 & bzq & 2.1677 & Massaro+14 & 5BZQ J1521+4336\\
J1548.4+1455 &  &  & J154824.38+145702.8 & bzg & 0.23 & Crespo+16b & \\
J1549.0+6309 & bcu II & SDSS J154958.45+631021.2 & J154828.42+631051.1 & bzg & 0.269 & Massaro+15b & 5BZG J1548+6310\\
J1549.5+1709 & bcu II & MG1 J154930+1708 & J154929.28+170828.1 & bzb & 0.0 & Crespo+16c & \\
J1553.5-3118 & bll & 1RXS J155333.4-311841 & J155333.56-311830.9 & bzb & 0.0 & Ricci+15 & 5BZB J1553-3118\\
J1611.9+1404 &  &  & J161118.10+140328.7 & qso & 0.5855 & Massaro+14 & \\
J1626.4-7640 & bcu II & PKS 1619-765 & J162638.15-763855.5 & bzb & 0.1050 & Ricci+15 & 5BZU J1626-7638\\
J1627.8+3217 &  &  & J162800.39+322414.0 & qso & 0.9051 & Massaro+14 & \\
J1636.7+2624 & bcu II & NVSS J163651+262657 & J163651.46+262656.7 & bzb & 0.0 & Crespo+16c & \\
J1647.4+4950 & bcu I & SBS 1646+499 & J164734.91+495000.5 & bzq & 0.049 & Crespo+16a & 5BZU J1647+4950\\
J1649.4+5238 & bll & 87GB 164812.2+524023 & J164924.98+523515.0 & bzb & 0.0 & Ricci+15 & 5BZB J1649+5235\\
J1656.8-2010 & bcu II & 1RXS J165655.0-201049 & J165655.14-201056.2 & bzb & 0.0 & Crespo+16c & \\
\hline 
\end{tabular}\\
Column description: 
 (1) 3FGL name reported; 
 (2) 3FGL $\gamma$-ray classification. Empty field implies that the sources is an UGS;  
 (3) 3FGL associated counterpart if present;
 (4) \wse\ name of the counterpart observed during our optical spectroscopic campaign;
 (5) optical classification
 (6) redshift; question marks indicate uncertain values or redshifts estimated via intervening systems while empty field is used for unknown $z$; 
 (7) reference within the papers published with the results of our campaign;
 (8) the name reported in the \bzcat.\\
Notes: Sources pointed twice are reported in different lines within the table.
\end{table*}

\begin{table*}
\tiny
 \caption{Summary of the targets observed during our optical spectroscopic campaign for the 3FGL sources. (R.A. 16-24)}
\label{tab:main3}
\begin{tabular}{|llllllll|}
\hline
3FGL    & 3FGL   & 3FGL & counterpart & class & $z$ & reference & \bzcat\ \\
name & class & counterpart & name          &           &        &                 & name \\
\hline
\hline 
J1704.1+1234 &  &  & J170409.58+123421.7 & bzb & 0.45 & Crespo+16b & \\
J1719.2+1744 & bll & PKS 1717+177 & J171913.05+174506.5 & bzb & 0.0 & Massaro+15b & 5BZB J1719+1745\\
J1725.3+5853 & bll & 7C 1724+5854 & J172535.02+585140.0 & bzb & 0.0 & Paggi+14 & 5BZB J1725+5851\\
J1730.6-0357 &  &  & J173052.85-035247.2 & bzb & 0.776? & Ricci+15 & \\
J1731.9+5428 &  &  & J173340.31+542636.7 & bzb & 0.0 & Massaro+16 & \\
J1736.0+2033 & bcu II & NVSS J173605+203301 & J173605.25+203301.1 & bzb & 0.0 & Crespo+16a & 5BZB J1736+2033\\
J1801.5+4403 & fsrq & S4 1800+44 & J180132.32+440421.7 & bzq & 0.663 & Massaro+15b & 5BZQ J1801+4404\\
J1809.7+2909 & bll & MG2 J180948+2910 & J180945.39+291019.8 & bzb & 0.0 & Ricci+15 & 5BZB J1809+2910\\
J1819.1+2134 & bcu II & MG2 J181902+2132 & J181905.22+213233.8 & bzb & 0.0 & Crespo+16b & \\
J1835.4+1349 & bcu III & TXS 1833+137 & J183535.34+134848.8 & bzb & 0.0 & Ricci+15 & \\
J1836.3+3137 & bll & RX J1836.2+3136 & J183621.23+313626.8 & bzb & 0.0 & Massaro+15a & 5BZB J1836+3136\\
J1844.3+1547 & bcu II & NVSS J184425+154646 & J184425.36+154645.8 & bzb & 0.0 & Massaro+15b & \\
J1844.1+5709 & bll & TXS 1843+571 & J184450.96+570938.6 & bzb & 0.0 & Massaro+15a & 5BZB J1844+5709\\
J1849.5+2751 & bll & MG2 J184929+2748 & J184931.74+274800.8 & bzb & 0.0 & Ricci+15 & 5BZB J1849+2748\\
J1903.2+5541 & bll & TXS 1902+556 & J190311.60+554038.5 & bzb & 0.0 & Massaro+15a & 5BZB J1903+5540\\
J1913.9+4441 & bcu II & 1RXS J191401.9+443849 & J191401.88+443832.2 & bzb & 0.0 & Crespo+16a & \\
J1942.7+1033 & bcu II & 1RXS J194246.3+103339 & J194247.48+103327.0 & bzb & 0.0 & Massaro+15b & 5BZB J1942+1033\\
J1955.0-1605 & bcu II & 1RXS J195500.6-160328 & J195500.65-160338.4 & bzb & 0.0 & Crespo+16c & \\
J1959.8-4725 & bcu II & SUMSS J195945-472519 & J195945.66-472519.3 & bzb & 0.519? & Ricci+15 & \\
J2004.8+7003 & bll & 1RXS J200504.0+700445 & J200505.97+700439.5 & bzb & 0.0 & Massaro+15a & 5BZB J2005+7004\\
J2014.5+0648 & bcu II & NVSS J201431+064849 & J201431.08+064852.5 & bzg & 0.341 & Massaro+15b & 5BZB J2014+0648\\
J2015.3-1431 &  &  & J201525.02-143203.9 & bzb & 0.0 & Crespo+16b & \\
J2021.9+0630 & bll & 87GB 201926.8+061922 & J202155.45+062913.6 & bzb & 0.0 & Crespo+16a & \\
J2031.0+1937 & bcu II & RX J2030.8+1935 & J203057.12+193612.9 & bzb & 0.668? & Massaro+15b & \\
J2036.4+6551 & bll & 87GB 203539.4+654245 & J203620.14+655314.5 & bzb & 0.0 & Crespo+16a & 5BZB J2036+6553\\
J2036.6-3325 & bcu II & 1RXS J203650.9-332817 & J203649.49-332830.7 & bzb & 0.23 & Crespo+16b & \\
J2039.5+5217 & bll & 1ES 2037+521 & J203923.51+521950.1 & bzb & 0.0 & Ricci+15 & 5BZB J2039+5219\\
J2104.2-0211 & bcu II & NVSS J210421-021239 & J210421.92-021239.0 & bzb & 0.0 & Crespo+16c & \\
J2108.0+3654 & bcu II & TXS 2106+367 & J210805.46+365526.5 & bzb & 0.0 & Massaro+15a & \\
J2109.1-6638 & bcu II & PKS 2104-668 & J210851.80-663722.7 & bzb & 0.0 & Crespo+16c & 5BZB J2108-6637\\
J2118.0-3241 & bcu I & NVSS J211754-324326 & J211754.91-324328.2 & bzb & 0.0 & Crespo+16c & \\
J2127.7+3612 & bll & B2 2125+35 & J212743.03+361305.7 & bzb & 0.0 & Massaro+15a & 5BZB J2127+3613\\
J2133.3+2533 & bcu II & 87GB 213100.1+251534 & J213314.36+252859.0 & bzg & 0.294 & Massaro+15b & 5BZG J2133+2528\\
J2134.5-2131 &  &  & J213430.18-213032.8 & bzb & 0.0 & Crespo+16b & \\
J2145.5+1007 &  &  & J214530.19+100605.5 & bzb & 0.0 & Massaro+16 & \\
J2156.0+1818 & bcu II & RX J2156.0+1818 & J215601.64+181837.1 & bzb & 0.0 & Crespo+16a & \\
J2223.3+0103 &  &  & J222329.57+010226.7 & bzb & 0.29? & Massaro+14 & 5BZB J2223+0102\\
J2223.3+0103 &  &  & J222329.57+010226.7 & bzb & 0.0 & Massaro+16 & 5BZB J2223+0102\\
J2232.9-2021 & bcu II  & 1RXS J223249.5-202232 & J223248.80-202226.2 & bzb & 0.0 & Crespo+16c & \\
J2247.8+4413 & bll & NVSS J224753+441317 & J224753.22+441315.5 & bzb & 0.0 & Massaro+15a & 5BZB J2247+4413\\
J2251.9+4031 & bll & MG4 J225201+4030 & J225159.77+403058.0 & bzb & 0.0 & Massaro+15a & 5BZB J2251+4030\\
J2255.1+2411 & bll & MG3 J225517+2409 & J225515.37+241011.2 & bzb & 0.0 & Massaro+15a & 5BZB J2255+2410\\
J2258.2-3645 &  &  & J225815.00-364434.3 & bzb & 0.0 & Landoni+15 & 5BZB J2258-3644\\
J2300.3+3136 & bll & NVSS J230022+313703 & J230022.84+313704.4 & bzb & 0.0 & Massaro+15b & 5BZB J2300+3137\\
J2311.0+0204 & bll & NVSS J231101+020504 & J231101.29+020505.3 & bzb & 0.0 & Massaro+15a & 5BZB J2311+0205\\
J2317.3-4534 & bcu II & 1RXS J231733.0-453348 & J231731.98-453359.6 & bzb & 0.0 & Crespo+16c & \\
J2323.9+4211 & bll & 1ES 2321+419 & J232352.07+421058.5 & bzb & 0.0 & Crespo+16a & 5BZB J2323+4210\\
J2323.9+4211 & bll & 1ES 2321+419 & J232352.07+421058.5 & bzb & 0.0 & Massaro+15a & 5BZB J2323+4210\\
J2324.7+0801 & bll & PMN J2324+0801 & J232445.32+080206.1 & bzb & 0.0 & Massaro+15a & 5BZB J2324+0802\\
J2325.6+1650 & bll & NVSS J232538+164641 & J232538.11+164642.7 & bzb & 0.0 & Massaro+15a & 5BZB J2325+1646\\
J2328.4-4034 & bcu II & PKS 2325-408 & J232819.26-403509.8 & bzq & 1.972 & Crespo+16c & 5BZQ J2328-4035\\
J2339.0+2122 & bll & RX J2338.8+2124 & J233856.38+212441.3 & bzb & 0.0 & Massaro+15a & 5BZB J2338+2124\\
J2340.7+8016 & bll & 1RXS J234051.4+801513 & J234054.23+801516.0 & bzb & 0.0 & Massaro+15b & 5BZB J2340+8015\\
J2346.7+0705 & bcu II & TXS 2344+068 & J234639.93+070506.8 & bzb & 0.17 & Crespo+16c & \\
J2352.0+1752 & bll & CLASS J2352+1749 & J235205.84+174913.7 & bzb & 0.0 & Massaro+15a & 5BZB J2352+1749\\
J2356.0+4037 & bll & NVSS J235612+403648 & J235612.70+403646.8 & bzg & 0.131 & Massaro+15a & 5BZG J2356+4036\\
\hline 
\end{tabular}\\
Column description: 
 (1) 3FGL name reported; 
 (2) 3FGL $\gamma$-ray classification. Empty field implies that the sources is an UGS;  
 (3) 3FGL associated counterpart if present;
 (4) \wse\ name of the counterpart observed during our optical spectroscopic campaign;
 (5) optical classification
 (6) redshift; question marks indicate uncertain values or redshifts estimated via intervening systems while empty field is used for unknown $z$; 
 (7) reference within the papers published with the results of our campaign;
 (8) the name reported in the \bzcat.\\
Notes: Sources pointed twice are reported in different lines within the table.
\end{table*}

\begin{table*}
\tiny
 \caption{Summary of the targets observed during our optical spectroscopic campaign for the 1FGL, 2FGL and 1FHL sources only.}
\label{tab:add}
\begin{tabular}{|llllllll|}
\hline
\fer\    & \fer\   & \fer\  & counterpart & class & $z$ & reference & \bzcat\ \\
name & class & counterpart & name          &           &        &                 & name \\
\hline
\hline 
1FHL J0030.1-1647 & & & J003020.44-164713.1 & bzb & 0.237 & Crespo+16b \\
1FHL J0044.0-1111 & & & J004348.66-111607.2 & bzb & 0.264 & Crespo+16b \\
2FGL J0332.5-1118 & agu & NVSS J033223-111951 & J033223.25-111950.6 & bzq & 0.2074 & Crespo+16b & 5BZU J0332-1119\\
1FGL J0411.6+5459 & & & J041203.78+545747.2 & bzb & 0.0 & Crespo+16b \\
2FGL J0819.6-0803 & bzb & RX J0819.2-0756 & J081917.58-075626.0 & bzb & 0.0 & Crespo+16b & 5BZB J0819-0756\\
1FGL J1422.7+3743 & bzb & CLASS J1423+3737 & J142304.61+373730.5 & bzb & 0.0 & Massaro+14 & 5BZB J1423+3737\\
2FGL J1624.4+1123 & agu & MG1 J162441+1111 & J162444.79+110959.3 & bzq & 2.1 & Crespo+16b \\
2FGL J1745.6+0203 & & & J174507.82+015442.4 & qso & 0.078 & Ricci+15 \\
2FGL J1803.6+2523c& agu & TXS 1801+253 & 2MASS J18031240+2521185 & bzq & 0.77 & Crespo+16b \\
2FGL J1848.6+3241 & agu & B2 1846+32B & J184834.37+324400.0 & bzq & 0.981 & Massaro+15b \\
2FGL J2110.3+3822 &  & & J211020.19+381659.2 & qso & 0.46 & Massaro+15a \\
\hline 
\end{tabular}\\
Column description: 
 (1) \fer\ name reported in the 1FGL, 2FGL or 1FHL; 
 (2) \fer\ $\gamma$-ray classification;  
 (3) \fer\ associated counterpart if present;
 (4) \wse\ or 2MASS name of the counterpart observed during our optical spectroscopic campaign;
 (5) optical classification
 (6) redshift; question marks indicate uncertain values or redshifts estimated via intervening systems while empty field is used for unknown $z$; 
 (7) reference within the papers published with the results of our campaign;
 (8) the name reported in the \bzcat.
\end{table*}

\end{document}